\documentclass[10.8pt]{article}%
\usepackage{amsfonts}
\usepackage{amsmath}
\usepackage{amssymb}
\usepackage{charter}
\usepackage{graphicx}
\usepackage{cite}
\usepackage{geometry}%
\providecommand{\U}[1]{\protect \rule{.1in}{.1in}}

\RequirePackage{CJK}
\AtBeginDocument{\begin{CJK*}{GBK}{song}\CJKtilde}
\AtEndDocument{\end{CJK*}}
\begin{document}

\title{Three-electron spin states and entanglement states}
\author{{\small Xiao-Jing Liu}$^{a,b}${\small , Xiang-Yao Wu}$^{b}$%
{\small \thanks{E-mail: wuxy2066@163.com}, Jing-Bin Lu}$^{a}${\small , Ji
Ma}$^{b}${\small , Hong-Li}$^{b}${\small , Si-Qi Zhang}$^{b}${\small , }
\and {\small Guang-Huai Wang}$^{b}${\small , Heng-Mei Li}$^{c}${\small , Hong-Chun
Yuan}$^{c}${\small  and Hai-Xin Gao}$^{d}${\small  }\\{\small a) Institute of Physics, Jilin University, Changchun 130012 China } \\{\small b) Institute of Physics, Jilin Normal University, Siping 136000 China} \\{\small c) School of Science, Changzhou Institute of Technology, Changzhou
213002, China}\\{\small d) Institute of Physics, Northeast Normal University, Changchun 130024
China} }
\maketitle

\begin{abstract}
In this paper, we have given the symmetrical and antisymmetrical spin and
space wave functions of three-electron, and further given the full total
entanglement states for the three-electron, which are related to their space
and spin wave function. When we study particles entanglement we not only
consider their spin entanglement and also consider their space entanglement.
Otherwise, we find that electrons entanglement are restricted to the teeny
range, when electrons exceed the space range, their entanglement should be
broken down even disappearance, which is accordance with the experiments results.



\end{abstract}

\section{Introduction}

The quantum many-body problem is one of the most fascinating topics in modern
physics, and as well as one of the most challenging\cite{s1}. The challenges
stem from the Hilbert space growing exponentially with the magnitude of the
systems if they are treated exactly. Fortunately, we do not need to address
the entire Hilbert space, because the physical properties of the quantum
many-body systems are determined by the ground state and some low excitation
levels \cite{s2}.

Entanglement is one of the most fundamentally nonclassical features of quantum
mechanics and as such is highly important to the foundations of modern
physics, e.g., in high energy physics \cite{s3,s4}. In a more practical view,
entanglement is most relevant in two distinct respects: technological
applications (such as e.g. quantum cryptography \cite{s5,s6} or future quantum
computers \cite{s7}).

In recent years, frustrated spin systems have received a lot of attention due
to its association with high-Tc super- conductivity \cite{s8}, and the
discovery of exotic frustrated phases \cite{s9}. There have been several
attempts to characterize and study frustrated systems using quantum
information concepts, which include entanglement area law \cite{s10,s11},
fidelity \cite{s12}, and quantum discord \cite{s13}. Various aspects of
entanglement have also been experimentally investigated in frustrated quantum
spin systems \cite{s14,s15}.

In this paper, we have given the symmetrical and antisymmetrical spin and
space wave functions of three-electron, and further given the full total
entanglement states for the three-electron, which are related to their space
and spin wave function. When we study particles entanglement we not only
consider their spin entanglement and also consider their space entanglement.
Otherwise, we find that electrons entanglement are restricted to the teeny
range, when electrons exceed the space range, their entanglement should be
broken down even disappearance, which is accordance with the experiments results.

\section{Two-electron spin states }

In quantum mechanics, the total spin $\vec{S}$ is the sum of three electrons
spin $\vec{s}_{1}$, $\vec{s}_{2}$ and $\vec{s}_{3}$
\begin{equation}
\vec{S}_{123}=\vec{s}_{1}+\vec{s}_{2}+\vec{s}_{3}=\vec{S}_{12}+\vec{s}%
_{3},\label{1}%
\end{equation}
where
\begin{equation}
\vec{S}_{12}=\vec{s}_{1}+\vec{s}_{2},\label{2}%
\end{equation}
and $\vec{S}_{12}^{2}$, $\vec{S}_{123}^{2}$ are
\begin{align}
\vec{S}_{12}^{2}  & =(\vec{s}_{1}+\vec{s}_{2})\hspace{0.001in}^{2}\nonumber \\
& =\vec{s}_{1}\hspace{0.001in}^{2}+\vec{s}_{2}\hspace{0.001in}^{2}+2\vec
{s}_{1}\cdot \vec{s}_{2}\nonumber \\
& =\frac{3}{2}+2\vec{s}_{1}\cdot \vec{s}_{2},\label{3}%
\end{align}

\begin{align}
\vec{S}_{123}^{2}  & =(\vec{s}_{1}+\vec{s}_{2}+\vec{s}_{3})\hspace
{0.001in}^{2}\nonumber \\
& =\vec{s}_{1}\hspace{0.001in}^{2}+\vec{s}_{2}\hspace{0.001in}^{2}+\vec{s}%
_{3}\hspace{0.001in}^{2}+2(\vec{s}_{1}\cdot \vec{s}_{2}+\vec{s}_{2}\cdot \vec
{s}_{3}+\vec{s}_{1}\cdot \vec{s}_{3})\nonumber \\
& =\frac{9}{4}+2(\vec{s}_{1}\cdot \vec{s}_{2}+\vec{s}_{2}\cdot \vec{s}_{3}%
+\vec{s}_{1}\cdot \vec{s}_{3}),\label{4}%
\end{align}
Obviously, $\vec{S}_{12}^{2}$ commutes with $\vec{S}_{12}$ and $\vec{s}_{3}$,
and $\vec{S}_{12}^{2}$ commutes with $\vec{S}_{123}$. So, $\vec{S}_{12}^{2}$,
$\vec{S}_{123}^{2}$ and $(\vec{S}_{123})_{z}$ commute with each other, and
they have common eigenfunctions. The eigenvalues and quantum number of
$\vec{S}_{12}^{2}$ and $\vec{S}_{123}^{2}$ are
\begin{equation}
\vec{S}_{12}^{2}=S^{\prime}(S^{\prime}+1),\hspace{0.3in}S^{\prime
}=0,1,\label{5}%
\end{equation}%
\begin{equation}
\vec{S}_{123}^{2}=S(S+1),\hspace{0.3in}S=\frac{1}{2}(S^{\prime}=0);\hspace
{0.2in}\frac{1}{2},\frac{3}{2}(S^{\prime}=1).\label{6}%
\end{equation}
For two-electron, $\vec{S}_{12}^{2}$ and $(\vec{S}_{12})_{z}$ have common spin
eigenstates, they are:\newline%
\begin{equation}
\chi_{11}(1,2)=\chi_{\frac{1}{2}}(s_{1z})\chi_{\frac{1}{2}}(s_{2z}%
),\hspace{0.3in}(S^{\prime}=1,\hspace{0.1in}(s_{12})_{z}=1)\label{7}%
\end{equation}

\begin{equation}
\chi_{10}(1,2)=\frac{1}{\sqrt{2}}[\chi_{\frac{1}{2}}(s_{1z})\chi_{-\frac{1}%
{2}}(s_{2z})+\chi_{\frac{1}{2}}(s_{2z})\chi_{-\frac{1}{2}}(s_{1z}%
)],\hspace{0.3in}(S^{\prime}=1,\hspace{0.1in}(s_{12})_{z}=0)\label{8}%
\end{equation}

\begin{equation}
\chi_{1-1}(1,2)=\chi_{-\frac{1}{2}}(s_{1z})\chi_{-\frac{1}{2}}(s_{2z}%
),\hspace{0.3in}(S^{\prime}=1,\hspace{0.1in}(s_{12})_{z}=-1)\label{9}%
\end{equation}

\begin{equation}
\chi_{00}(1,2)=\frac{1}{\sqrt{2}}[\chi_{\frac{1}{2}}(s_{1z})\chi_{-\frac{1}%
{2}}(s_{2z})-\chi_{\frac{1}{2}}(s_{2z})\chi_{-\frac{1}{2}}(s_{1z}%
)],\hspace{0.3in}(S^{\prime}=0,\hspace{0.1in}(s_{12})_{z}=0)\label{10}%
\end{equation}
where $\chi_{\frac{1}{2}}(s_{1z})$ and $\chi_{\frac{1}{2}}(s_{2z})$ are the
spin eigenstates of $s_{1z}$ and $s_{2z}$.\newline Defining single-electron
lowering operators $s_{i-}$ as
\begin{equation}
s_{i-}=s_{ix}-is_{iy},\hspace{0.2in}(i=1,2,3)\label{11}%
\end{equation}
and the total lowering operator of three-electron is
\begin{equation}
S_{123-}=s_{1-}+s_{2-}+s_{3-},\label{12}%
\end{equation}
by the angular theory, we have
\begin{equation}
s_{-}|jm\rangle=\sqrt{(j+m)(j-m+1)}|jm-1\rangle.\label{13}%
\end{equation}
By equation (\ref{13}), we have
\begin{equation}
s_{1-}\chi_{\frac{1}{2}}(s_{1z})=\chi_{-\frac{1}{2}}(s_{1z}),\hspace
{0.3in}s_{1-}\chi_{-\frac{1}{2}}(s_{1z})=0\label{14}%
\end{equation}

\begin{equation}
s_{2-}\chi_{\frac{1}{2}}(s_{2z})=\chi_{-\frac{1}{2}}(s_{2z}),\hspace
{0.3in}s_{2-}\chi_{-\frac{1}{2}}(s_{2z})=0\label{15}%
\end{equation}

\begin{equation}
s_{3-}\chi_{\frac{1}{2}}(s_{3z})=\chi_{-\frac{1}{2}}(s_{3z}),\hspace
{0.3in}s_{3-}\chi_{-\frac{1}{2}}(s_{3z})=0.\label{16}%
\end{equation}

\section{Three-electron spin states}

In the following, we should give the spin states of three-electron. For
two-electron , the states of $S^{\prime}=1$ are symmetrical. So, all the
three-electron states of $S^{\prime}=1$ should be symmetrical. For
two-electron, the states of $S^{\prime}=0$ are antisymmetrical. So, all the
three-electron states of $S^{\prime}=0$ should be antisymmetrical. We should
give all symmetrical and antisymmetrical states for three-electron.\newline(1)
$S\hspace{0.02in}^{\prime}=1$, $S=\frac{3}{2}$ spin wave functions
$\chi_{S^{\prime}SM}$ of three-electron\newline(a) The spin wave function
$\chi_{1\frac{3}{2}\frac{3}{2}}$\newline the three quantum number take maximum
value, i.e., $S^{\prime}=1$, $S=\frac{3}{2}$ and $M=\frac{3}{2}$. It can be
written directly
\begin{align}
\chi_{1\frac{3}{2}\frac{3}{2}}  & =\chi_{11}(1,2)\chi_{\frac{1}{2}}%
(s_{3z})\nonumber \\
& =\chi_{\frac{1}{2}}(s_{1z})\chi_{\frac{1}{2}}(s_{2z})\chi_{\frac{1}{2}%
}(s_{3z}).\label{17}%
\end{align}
The state $\chi_{1\frac{3}{2}\frac{3}{2}}$ is symmetrical when exchange
electron $1$, $2$ and $3$.\newline(b) The spin wave function $\chi_{1\frac
{3}{2}\frac{1}{2}}$\newline By operating lowering operators $S_{123-}$ on
$\chi_{1\frac{3}{2}\frac{3}{2}}$, we can obtain $\chi_{1\frac{3}{2}\frac{1}%
{2}}$
\begin{align}
S_{123-}\chi_{1\frac{3}{2}\frac{3}{2}}  & =(s_{1-}+s_{2-}+s_{3-})\chi
_{\frac{1}{2}}(s_{1z})\chi_{\frac{1}{2}}(s_{2z})\chi_{\frac{1}{2}}%
(s_{3z})\nonumber \\
& =(s_{1-}\chi_{\frac{1}{2}}(s_{1z}))\chi_{\frac{1}{2}}(s_{2z})\chi_{\frac
{1}{2}}(s_{3z})+\chi_{\frac{1}{2}}(s_{1z})(s_{2-}\chi_{\frac{1}{2}}%
(s_{2z}))\chi_{\frac{1}{2}}(s_{3z})\nonumber \\
& +\chi_{\frac{1}{2}}(s_{1z})\chi_{\frac{1}{2}}(s_{2z})(s_{3-}\chi_{\frac
{1}{2}}(s_{3z}))\nonumber \\
& =\chi_{-\frac{1}{2}}(s_{1z})\chi_{\frac{1}{2}}(s_{2z})\chi_{\frac{1}{2}%
}(s_{3z})+\chi_{\frac{1}{2}}(s_{1z})\chi_{-\frac{1}{2}}(s_{2z})\chi_{\frac
{1}{2}}(s_{3z})\nonumber \\
& +\chi_{\frac{1}{2}}(s_{1z})\chi_{\frac{1}{2}}(s_{2z})\chi_{-\frac{1}{2}%
}(s_{3z}),\label{18}%
\end{align}
by equation (\ref{13}), we have
\begin{equation}
S_{123-}\chi_{1\frac{3}{2}\frac{3}{2}}=\sqrt{3}\chi_{1\frac{3}{2}\frac{1}{2}%
},\label{19}%
\end{equation}
or
\begin{align}
\chi_{1\frac{3}{2}\frac{1}{2}}  & =\frac{1}{\sqrt{3}}[\chi_{-\frac{1}{2}%
}(s_{1z})\chi_{\frac{1}{2}}(s_{2z})\chi_{\frac{1}{2}}(s_{3z})+\chi_{\frac
{1}{2}}(s_{1z})\chi_{-\frac{1}{2}}(s_{2z})\chi_{\frac{1}{2}}(s_{3z}%
)\nonumber \\
& +\chi_{-\frac{1}{2}}(s_{1z})\chi_{\frac{1}{2}}(s_{2z})\chi_{-\frac{1}{2}%
}(s_{3z})].\label{20}%
\end{align}
The state $\chi_{1\frac{3}{2}\frac{1}{2}}$ is symmetrical when exchange
electron $1$, $2$ and $3$.\newline(c) The spin wave function $\chi_{1\frac
{3}{2}-\frac{1}{2}}$\newline By operating lowering operators $S_{123-}$ on
$\chi_{1\frac{3}{2}\frac{1}{2}}$, we can obtain
\begin{align}
S_{123-}\chi_{1\frac{3}{2}\frac{1}{2}}  & =\frac{1}{\sqrt{3}}[(s_{1-}%
\chi_{-\frac{1}{2}}(s_{1z}))\chi_{\frac{1}{2}}(s_{2z})\chi_{\frac{1}{2}%
}(s_{3z})+(s_{1-}\chi_{\frac{1}{2}}(s_{1z}))\chi_{-\frac{1}{2}}(s_{2z}%
)\chi_{\frac{1}{2}}(s_{3z})\nonumber \\
& +(s_{1-}\chi_{\frac{1}{2}}(s_{1z}))\chi_{\frac{1}{2}}(s_{2z})\chi_{-\frac
{1}{2}}(s_{3z})+\chi_{-\frac{1}{2}}(s_{1z})(s_{2-}\chi_{\frac{1}{2}}%
(s_{2z}))\chi_{\frac{1}{2}}(s_{3z})\nonumber \\
& +\chi_{\frac{1}{2}}(s_{1z})(s_{2-}\chi_{-\frac{1}{2}}(s_{2z}))\chi_{\frac
{1}{2}}(s_{3z})+\chi_{\frac{1}{2}}(s_{1z})(s_{2-}\chi_{\frac{1}{2}}%
(s_{2z}))\chi_{-\frac{1}{2}}(s_{3z})\nonumber \\
& +\chi_{-\frac{1}{2}}(s_{1z})\chi_{\frac{1}{2}}(s_{2z})(s_{3-}\chi_{\frac
{1}{2}}(s_{3z}))+\chi_{\frac{1}{2}}(s_{1z})\chi_{-\frac{1}{2}}(s_{2z}%
)(s_{3-}\chi_{\frac{1}{2}}(s_{3z}))\nonumber \\
& +\chi_{\frac{1}{2}}(s_{1z})\chi_{\frac{1}{2}}(s_{2z})(s_{3-}\chi_{-\frac
{1}{2}}(s_{3z}))\nonumber \\
& =\frac{1}{\sqrt{3}}[\chi_{-\frac{1}{2}}(s_{1z})\chi_{-\frac{1}{2}}%
(s_{2z})\chi_{\frac{1}{2}}(s_{3z})+\chi_{-\frac{1}{2}}(s_{1z})\chi_{\frac
{1}{2}}(s_{2z})\chi_{-\frac{1}{2}}(s_{3z})\nonumber \\
& +\chi_{-\frac{1}{2}}(s_{1z})\chi_{-\frac{1}{2}}(s_{2z})\chi_{\frac{1}{2}%
}(s_{3z})+\chi_{\frac{1}{2}}(s_{1z})\chi_{-\frac{1}{2}}(s_{2z})\chi_{-\frac
{1}{2}}(s_{3z})\nonumber \\
& +\chi_{-\frac{1}{2}}(s_{1z})\chi_{\frac{1}{2}}(s_{2z})\chi_{-\frac{1}{2}%
}(s_{3z})+\chi_{\frac{1}{2}}(s_{1z})\chi_{-\frac{1}{2}}(s_{2z})\chi_{-\frac
{1}{2}}(s_{3z})]\nonumber \\
& =\frac{2}{\sqrt{3}}[\chi_{-\frac{1}{2}}(s_{1z})\chi_{-\frac{1}{2}}%
(s_{2z})\chi_{\frac{1}{2}}(s_{3z})+\chi_{-\frac{1}{2}}(s_{1z})\chi_{\frac
{1}{2}}(s_{2z})\chi_{-\frac{1}{2}}(s_{3z})\nonumber \\
& +\chi_{\frac{1}{2}}(s_{1z})\chi_{-\frac{1}{2}}(s_{2z})\chi_{-\frac{1}{2}%
}(s_{3z})],\label{21}%
\end{align}
by equation (\ref{13}), we have
\begin{equation}
S_{123-}\chi_{1\frac{3}{2}\frac{1}{2}}=\sqrt{4}\chi_{1\frac{3}{2}-\frac{1}{2}%
},\label{22}%
\end{equation}
or
\begin{align}
\chi_{1\frac{3}{2}-\frac{1}{2}}  & =\frac{1}{\sqrt{3}}[\chi_{-\frac{1}{2}%
}(s_{1z})\chi_{-\frac{1}{2}}(s_{2z})\chi_{\frac{1}{2}}(s_{3z})+\chi_{-\frac
{1}{2}}(s_{1z})\chi_{\frac{1}{2}}(s_{2z})\chi_{-\frac{1}{2}}(s_{3z}%
)\nonumber \\
& +\chi_{\frac{1}{2}}(s_{1z})\chi_{-\frac{1}{2}}(s_{2z})\chi_{-\frac{1}{2}%
}(s_{3z})].\label{23}%
\end{align}
The state $\chi_{1\frac{3}{2}-\frac{1}{2}}$ is symmetrical when exchange
electron $1$, $2$ and $3$.\newline(d) The spin wave function $\chi_{1\frac
{3}{2}-\frac{3}{2}}$\newline when quantum number $M$ take minimum value, i.e.,
$M=-\frac{3}{2}$, the spin wave function of three-electron can be written
directly
\begin{equation}
\chi_{1\frac{3}{2}-\frac{3}{2}}=\chi_{-\frac{1}{2}}(s_{1z})\chi_{-\frac{1}{2}%
}(s_{2z})\chi_{-\frac{1}{2}}(s_{3z}).\label{24}%
\end{equation}
The state $\chi_{1\frac{3}{2}-\frac{3}{2}}$ is symmetrical when exchange
electron $1$, $2$ and $3$.\newline%
\begin{equation}
\chi_{1\frac{3}{2}\frac{1}{2}}=\frac{1}{\sqrt{3}}[\chi_{11}(1,2)\chi
_{-\frac{1}{2}}(3)+\sqrt{2}\chi_{10}(1,2)\chi_{\frac{1}{2}}(3)],\label{25}%
\end{equation}
(2) $S\hspace{0.02in}^{\prime}=1$, $S=\frac{1}{2}$ spin wave functions
$\chi_{S^{\prime}SM}$ of three-electron\newline(a) The spin wave function
$\chi_{1\frac{1}{2}\frac{1}{2}}$\newline We firstly calculate $\chi_{1\frac
{1}{2}\frac{1}{2}}$ state, it includes two $\chi_{\frac{1}{2}}$ states and one
$\chi_{-\frac{1}{2}}$ state, and it is the linear superposition of $\chi
_{11}(s_{1z},s_{2z})\chi_{-\frac{1}{2}}(s_{3z})$ and $\chi_{10}(s_{1z}%
,s_{2z})\chi_{\frac{1}{2}}(s_{3z})$. Since $\chi_{1\frac{1}{2}\frac{1}{2}}$
and $\chi_{1\frac{3}{2}\frac{1}{2}}$ is orthogonal, and the $\chi_{1\frac
{3}{2}\frac{1}{2}}$ state can be written as
\begin{equation}
\chi_{1\frac{3}{2}\frac{1}{2}}=\frac{1}{\sqrt{3}}[\chi_{11}(1,2)\chi
_{-\frac{1}{2}}(3)+\sqrt{2}\chi_{10}(1,2)\chi_{\frac{1}{2}}(3)],\label{26}%
\end{equation}
and state $\chi_{1\frac{1}{2}\frac{1}{2}}$ can be written as
\begin{align}
\chi_{1\frac{1}{2}\frac{1}{2}}  & =\frac{1}{\sqrt{3}}[\sqrt{2}\chi
_{11}(1,2)\chi_{-\frac{1}{2}}(3)-\chi_{10}(1,2)\chi_{\frac{1}{2}%
}(3)]\nonumber \\
& =\frac{1}{\sqrt{6}}[2\chi_{\frac{1}{2}}(s_{1z})\chi_{\frac{1}{2}}%
(s_{2z})\chi_{-\frac{1}{2}}(s_{3z})-\chi_{\frac{1}{2}}(s_{1z})\chi_{-\frac
{1}{2}}(s_{2z})\chi_{\frac{1}{2}}(s_{3z})\nonumber \\
& -\chi_{-\frac{1}{2}}(s_{1z})\chi_{\frac{1}{2}}(s_{2z})\chi_{\frac{1}{2}%
}(s_{3z})]\nonumber \\
& =\chi_{1\frac{1}{2}\frac{1}{2}}^{S(12)}(123).\label{27}%
\end{align}
The state $\chi_{1\frac{1}{2}\frac{1}{2}}$ should be symmetrical, but equation
\textbf{(}\ref{27}\textbf{)} is symmetrical for exchanging photon $1$ and $2$,
and not full symmetrical form. It should be symmetrization, and it is
\begin{equation}
\chi_{1\frac{1}{2}\frac{1}{2}}^{S}=\frac{1}{\sqrt{3}}[\chi_{1\frac{1}{2}%
\frac{1}{2}}^{S(12)}(123)+\chi_{1\frac{1}{2}\frac{1}{2}}^{S(13)}%
(123)+\chi_{1\frac{1}{2}\frac{1}{2}}^{S(23)}(123)].\label{28}%
\end{equation}
where $\chi_{1\frac{1}{2}\frac{1}{2}}^{S(12)}{222}(123)$ is symmetrical for
exchanging photon $1$ and $2$, $\chi_{1\frac{1}{2}\frac{1}{2}}^{S(13)}(123)$
is symmetrical for exchanging photon $1$ and $3$, and the state $\chi
_{1\frac{1}{2}\frac{1}{2}}^{S(13)}(123)$ can be obtained by the state
$\chi_{1\frac{1}{2}\frac{1}{2}}^{S(12)}(123)$ exchanging photon $2$ and $3$.
Obviously, the state $\chi_{1\frac{1}{2}\frac{1}{2}}^{S}$ is full symmetrical
for exchanging photon $1$, $2$ and $3$.\newline(b) The spin wave function
$\chi_{1\frac{1}{2}-\frac{1}{2}}$\newline Since state $\chi_{1\frac{3}%
{2}-\frac{1}{2}}$ and state $\chi_{1\frac{1}{2}-\frac{1}{2}}$ is orthogonal,
and state $\chi_{1\frac{3}{2}-\frac{1}{2}}$ can be written as
\begin{equation}
\chi_{1\frac{3}{2}-\frac{1}{2}}=\frac{1}{\sqrt{3}}[\sqrt{2}\chi_{10}%
(1,2)\chi_{-\frac{1}{2}}(3)+\chi_{1-1}(1,2)\chi_{\frac{1}{2}}(3)],\label{29}%
\end{equation}
and state $\chi_{1\frac{1}{2}-\frac{1}{2}}$ can be written as
\begin{align}
\chi_{1\frac{1}{2}-\frac{1}{2}}  & =\frac{1}{\sqrt{3}}[\chi_{10}%
(1,2)\chi_{-\frac{1}{2}}(3)-\sqrt{2}\chi_{1-1}(1,2)\chi_{\frac{1}{2}%
}(3)]\nonumber \\
& =\frac{1}{\sqrt{6}}[\chi_{\frac{1}{2}}(1)\chi_{-\frac{1}{2}}(2)\chi
_{-\frac{1}{2}}(3)+\chi_{-\frac{1}{2}}(1)\chi_{\frac{1}{2}}(2)\chi_{-\frac
{1}{2}}(3)\nonumber \\
& -2\chi_{-\frac{1}{2}}(1)\chi_{-\frac{1}{2}}(2)\chi_{\frac{1}{2}%
}(3)]\nonumber \\
& =\chi_{1\frac{1}{2}-\frac{1}{2}}^{S(12)}(123).\label{30}%
\end{align}
The state $\chi_{1\frac{1}{2}-\frac{1}{2}}$ should be symmetrical, but
equation \textbf{(\ref{30})} is symmetrical for exchanging photon $1$ and $2$,
and not full symmetrical form. It should be symmetrization, and it is
\begin{equation}
\chi_{1\frac{1}{2}-\frac{1}{2}}^{S}=\frac{1}{\sqrt{3}}[\chi_{1\frac{1}%
{2}-\frac{1}{2}}^{S(12)}(123)+\chi_{1\frac{1}{2}-\frac{1}{2}}^{S(13)}%
(123)+\chi_{1\frac{1}{2}-\frac{1}{2}}^{S(23)}(123)].\label{31}%
\end{equation}
(3) $S\hspace{0.02in}^{\prime}=0$, $S=\frac{1}{2}$ spin wave functions
$\chi_{S^{\prime}SM}$ of three-electron\newline(a) The spin wave function
$\chi_{0\frac{1}{2}\frac{1}{2}}$\newline For $S\hspace{0.02in}^{\prime}=0$,
the spin wave functions of electron $1$ and $2$ is $\chi_{00}(1,2)$. When
$M=\frac{1}{2}$, the state $\chi_{0\frac{1}{2}\frac{1}{2}}$ can be written as
directly
\begin{align}
\chi_{0\frac{1}{2}\frac{1}{2}}  & =\chi_{00}(1,2)\chi_{\frac{1}{2}%
}(3)\nonumber \\
& =\frac{1}{\sqrt{2}}[\chi_{\frac{1}{2}}(1)\chi_{-\frac{1}{2}}(2)\chi
_{\frac{1}{2}}(3)-\chi_{-\frac{1}{2}}(1)\chi_{\frac{1}{2}}(2)\chi_{\frac{1}%
{2}}(3)]\nonumber \\
& =\chi_{0\frac{1}{2}\frac{1}{2}}^{A(12)}(123).\label{32}%
\end{align}
The state $\chi_{0\frac{1}{2}\frac{1}{2}}$ should be antisymmetrical, but
equation (\ref{32}) is antisymmetrical for exchanging photon $1$ and $2$, and
not full antisymmetrical form. It should be antisymmetrization, and it is
\begin{equation}
\chi_{0\frac{1}{2}\frac{1}{2}}^{A}=\frac{1}{\sqrt{3}}[\chi_{0\frac{1}{2}%
\frac{1}{2}}^{A(12)}(123)+\chi_{0\frac{1}{2}\frac{1}{2}}^{A(13)}%
(123)+\chi_{0\frac{1}{2}\frac{1}{2}}^{A(23)}(123)].\label{33}%
\end{equation}
(b) The spin wave function $\chi_{0\frac{1}{2}-\frac{1}{2}}$\newline When
$M=\frac{1}{2}$, the state $\chi_{0\frac{1}{2}-\frac{1}{2}}$ can be also
written as directly
\begin{align}
\chi_{0\frac{1}{2}-\frac{1}{2}}  & =\chi_{00}(1,2)\chi_{-\frac{1}{2}%
}(3)\nonumber \\
& =\frac{1}{\sqrt{2}}[\chi_{\frac{1}{2}}(1)\chi_{-\frac{1}{2}}(2)\chi
_{-\frac{1}{2}}(3)-\chi_{-\frac{1}{2}}(1)\chi_{\frac{1}{2}}(2)\chi_{-\frac
{1}{2}}(3)],\label{34}%
\end{align}
The state $\chi_{0\frac{1}{2}-\frac{1}{2}}$ should be antisymmetrical, but
equation (\ref{34}) is antisymmetrical for exchanging photon $1$ and $2$, and
not full antisymmetrical form. It should be antisymmetrization, and it is
\begin{equation}
\chi_{0\frac{1}{2}-\frac{1}{2}}^{A}=\frac{1}{\sqrt{3}}[\chi_{0\frac{1}%
{2}-\frac{1}{2}}^{A(12)}(123)+\chi_{0\frac{1}{2}-\frac{1}{2}}^{A(13)}%
(123)+\chi_{0\frac{1}{2}-\frac{1}{2}}^{A(23)}(123)].\label{35}%
\end{equation}

\section{Three-electron space states}

When we neglect the interaction among electrons, the space wave functions of
three-electron can be written as\newline(1) The symmetrical space wave
functions of three-electron
\begin{equation}
\psi_{1}^{S}(\vec{r}_{1},\vec{r}_{2},\vec{r}_{3})=\psi_{n}(\vec{r}_{1}%
)\psi_{n}(\vec{r}_{2})\psi_{n}(\vec{r}_{3}),\hspace{0.3in}(n=m=l)\label{36}%
\end{equation}%
\begin{align}
\psi_{2}^{S}(\vec{r}_{1},\vec{r}_{2},\vec{r}_{3})  & =\frac{1}{\sqrt{6}}%
[\psi_{n}(\vec{r}_{1})\psi_{m}(\vec{r}_{2})\psi_{l}(\vec{r}_{3})+\psi_{n}%
(\vec{r}_{1})\psi_{l}(\vec{r}_{2})\psi_{m}(\vec{r}_{3})\nonumber \\
& +\psi_{l}(\vec{r}_{1})\psi_{m}(\vec{r}_{2})\psi_{n}(\vec{r}_{3})+\psi
_{m}(\vec{r}_{1})\psi_{n}(\vec{r}_{2})\psi_{l}(\vec{r}_{3})\nonumber \\
& +\psi_{l}(\vec{r}_{1})\psi_{n}(\vec{r}_{2})\psi_{m}(\vec{r}_{3})+\psi
_{m}(\vec{r}_{1})\psi_{l}(\vec{r}_{2})\psi_{n}(\vec{r}_{3})],\hspace
{0.3in}(n\neq m\neq l)\label{37}%
\end{align}%
\begin{align}
\psi_{3}^{S}(\vec{r}_{1},\vec{r}_{2},\vec{r}_{3})  & =\frac{1}{\sqrt{3}}%
[\psi_{n}(\vec{r}_{1})\psi_{n}(\vec{r}_{2})\psi_{l}(\vec{r}_{3})+\psi_{n}%
(\vec{r}_{1})\psi_{m}(\vec{r}_{2})\psi_{n}(\vec{r}_{3})\nonumber \\
& +\psi_{l}(\vec{r}_{1})\psi_{n}(\vec{r}_{2})\psi_{n}(\vec{r}_{3}%
)],\hspace{0.3in}(n=m\neq l)\label{38}%
\end{align}
(2) The antisymmetrical space wave functions of three-electron
\begin{equation}
\psi_{1}^{A}(\vec{r}_{1},\vec{r}_{2},\vec{r}_{3})=\frac{1}{\sqrt{3!}%
}\left \vert
\begin{array}
[c]{ccc}%
\psi_{n}(\vec{r}_{1}) & \psi_{n}(\vec{r}_{2}) & \psi_{n}(\vec{r}_{3})\\
\psi_{m}(\vec{r}_{1}) & \psi_{m}(\vec{r}_{2}) & \psi_{m}(\vec{r}_{3})\\
\psi_{l}(\vec{r}_{1}) & \psi_{l}(\vec{r}_{2}) & \psi_{l}(\vec{r}_{3})
\end{array}
\right \vert .\label{39}%
\end{equation}
Where $\psi_{m}(\vec{r}_{1}$, $\psi_{n}(\vec{r}_{2})$ and $\psi_{l}(\vec
{r}_{3})$ are single electron wave function, and $m$, $n$ and $l$ express
quantum numbers of quantum state.

\section{Three-electron total wave functions and entanglement states}

The indistinguishability principle applied for fermions leads to the
multi-fermion total wave functions antisymmetry, i.e., the space wave function
is symmetrical, the spin wave function is antisymmetrical, or the space wave
function is antisymmetrical, the spin wave function is symmetrical. The
three-electron total wave functions are\newline(a) When $m=n=l$, the
three-electron total antisymmetrical wave function
\begin{equation}
\Psi_{1}^{A}(q_{1},q_{2},q_{3})=\psi_{1}^{S}(r_{1},r_{2},r_{3})\chi
_{S^{\prime}SM}^{A}(s_{1z},s_{2z},s_{3z}),\label{40}%
\end{equation}
(b) When $m\neq n\neq l$, the three-electron total antisymmetrical wave
function
\begin{equation}
\Psi_{2}^{A}(q_{1},q_{2},q_{3})=\psi_{2}^{S}(r_{1},r_{2},r_{3})\chi
_{S^{\prime}SM}^{A}(s_{1z},s_{2z},s_{3z}),\label{41}%
\end{equation}
(c) When $m=n\neq l$, the three-electron total antisymmetrical wave function
\begin{align}
\Psi_{3}^{A}(q_{1},q_{2},q_{3})  & =\frac{1}{\sqrt{3}}[\psi_{n}(\vec{r}%
_{1})\psi_{n}(\vec{r}_{2})\chi^{A}(1,2)\psi_{l}(\vec{r}_{3})\chi(3)+\psi
_{n}(\vec{r}_{1})\psi_{n}(\vec{r}_{3})\chi^{A}(1,3)\psi_{l}(\vec{r}_{2}%
)\chi(2))\nonumber \\
& +\psi_{n}(\vec{r}_{2})\psi_{n}(\vec{r}_{3})\chi^{A}(2,3)\psi_{l}(\vec{r}%
_{1})\chi(1)],\label{42}%
\end{align}
(d) When $m\neq n\neq l$, the three-electron total antisymmetrical wave
function
\begin{equation}
\Psi^{A}(q_{1},q_{2},q_{3})=\psi^{A}(r_{1},r_{2},r_{3})\chi_{S^{\prime}SM}%
^{S}(s_{1z},s_{2z},s_{3z}),\label{43}%
\end{equation}
where $q_{1}$, $q_{2}$ and $q_{3}$ express the space coordinate and spin
component of electron $1$, $2$ and $3$, respectively, the wave functions
$\psi^{S}(r_{1},r_{2},r_{3})$ and $\psi^{A}(r_{1},r_{2},r_{3})$ are
symmetrical and antisymmetrical space wave functions of three-electron, which
are shown in equations (\ref{36})-(\ref{39}), the wave functions
$\chi_{S^{\prime}SM}^{A}(s_{1z},s_{2z},s_{3z})$ and $\chi_{S^{\prime}SM}%
^{S}(s_{1z},s_{2z},s_{3z})$ are symmetrical and antisymmetrical spin wave
functions of three-electron, which are shown in section $3$. The equations
(\ref{40})-(\ref{43}) are the total wave functions of three-electron, i.e.,
three-electron full entanglement states, and we can obtain the results: (1)
The three-electron entanglement states is related to their space and spin wave
functions, and we should not only consider their spin wave function parts. (2)
when there are at least one entanglement state for the space wave function or
spin wave function, the total wave functions of three-electron are
entanglement state. (3) when the space wave function and spin wave function
are not entanglement state, the total wave functions of three-electron is not
on entanglement state. (4) When the space wave function is entanglement, and
the spin wave function is not entanglement, the three-electron is called the
space entanglement. (5) When the spin wave function is entanglement, and the
space wave function is not entanglement, the three-electron is called the spin
entanglement. (6) When the space and spin wave function are all entanglement,
the three-electron is called the full entanglement. (7) When the space and
spin wave function are not entanglement, the three-electron is not entanglement.

When three electrons are on bound state, whether they are in well potential or
central force field and so on, every electron space wave function $\psi
_{m}(\vec{r})\rightarrow0$ when $r\geq a_{0}$ ($a_{0}$ is a teeny number). The
symmetrical and antisymmetrical space wave functions $\Psi^{S}(q_{1}%
,q_{2},q_{3})$ and $\Psi^{A}(q_{1},q_{2},q_{3})$ of three-electron should be
tended to zero, then three-electron space entanglement should be broken down
and become without entanglement for the three-electron. In experiments
\cite{s16,s17}, the authors have found multiparticle restricted to one
entanglement bit rather than an arbitrary amount of entanglement, which are
agreement with our theory analysis results.

\section{Conclusion}

In this paper, we have given the symmetrical and antisymmetrical spin and
space wave functions of three-electron, and further given the full total
entanglement states for the three-electron, which are related to their space
and spin wave function. When we study particles entanglement we not only
consider their spin entanglement and also consider their space entanglement.
Otherwise, we find that electrons entanglement are restricted to the teeny
range, when electrons exceed the space range, their entanglement should be
broken down even disappearance, which is accordance with the experiments results.

\end{document}